\documentclass[prl,aps,twocolumn,superscriptaddress,showpacs]{revtex4-1}
\usepackage[colorlinks=true,linkcolor=black, citecolor=blue, urlcolor=blue, 
    unicode=true]{hyperref}

\usepackage{graphicx}
\usepackage{epstopdf}

\begin{document}

\title{Spin-wave directional anisotropies in antiferromagnetic Ba$_{3}$NbFe$_{3}$Si$_{2}$O$_{14}$}

\author{C. Stock}
\affiliation{School of Physics and Astronomy, University of Edinburgh, Edinburgh EH9 3JZ, UK}

\author{R. D. Johnson}
\affiliation{Oxford Physics, Clarendon Laboratory, Parks Road, Oxford OX1 3PU, UK}

\author{N. Giles-Donovan}
\affiliation{Medical and Industrial Ultrasonics, School of Engineering, University of Glasgow G128QQ, UK}

\author{M. Songvilay}
\affiliation{School of Physics and Astronomy, University of Edinburgh, Edinburgh EH9 3JZ, UK}

\author{J. A. Rodriguez-Rivera}
\affiliation{NIST Center for Neutron Research, National Institute of Standards and Technology, 100 Bureau Drive, Gaithersburg, Maryland, 20899, USA}
\affiliation{Department of Materials Science, University of Maryland, College Park, Maryland 20742, USA}

\author{N. Lee}
\author{X. Xu}
\affiliation{Rutgers Center for Emergent Materials and Department of Physics and Astronomy, Rutgers University, 136 Frelinghuysen Road, Piscataway, New Jersey 08854, USA}

\author{P.G. Radaelli}
\affiliation{Oxford Physics, Clarendon Laboratory, Parks Road, Oxford OX1 3PU, UK}

\author{L.C. Chapon}
\author{A. Bombardi}
\affiliation{Diamond Light Source Ltd., Harwell Science and Innovation Campus, Didcot, Oxfordshire OX11 0DE, United Kingdom}

\author{S. Cochran}
\affiliation{Medical and Industrial Ultrasonics, School of Engineering, University of Glasgow G128QQ, UK}

\author{Ch. Niedermayer}
\affiliation{Laboratory for Neutron Scattering, Paul Scherrer Institut, CH-5232 Villigen, Switzerland}

\author{A. Schneidewind}
\affiliation{Forschungsneutronenquell Heinz Meier-Leibnitz (FRM-II), D-85747 Garching, Germany}
\affiliation{Institut fur Festkorperphysiki, TU Dresden, D-1062 Dresden, Germany}

\author{Z. Husges}
\author{Z. Lu}
\author{S. Meng}
\affiliation{Helmholtz-Zentrum Berlin für Materialien und Energie, 14109 Berlin, Germany}

\author{S. -W. Cheong}
\affiliation{Rutgers Center for Emergent Materials and Department of Physics and Astronomy, Rutgers University, 136 Frelinghuysen Road, Piscataway, New Jersey 08854, USA}

\date{\today}

\begin{abstract}

Ba$_{3}$NbFe$_{3}$Si$_{2}$O$_{14}$ (langasite) is structurally and magnetically single domain chiral with the magnetic helicity induced through competing symmetric exchange interactions.  Using neutron scattering, we show that the spin-waves in antiferromagnetic langasite display directional anisotropy.  On applying a time reversal symmetry breaking magnetic field along the $c$-axis, the spin wave energies differ when the sign is reversed for either the momentum transfer $\pm$ $\vec{Q}$ or applied magnetic field $\pm$ $\mu_{0}$H.  When the field is applied within the crystallographic $ab$-plane, the spin wave dispersion is directionally \textit{isotropic} and symmetric in $\pm$ $\mu_{0}$H.  However, a directional anisotropy is observed in the spin wave intensity.  We discuss this directional anisotropy in the dispersion in langasite in terms of a field induced precession of the dynamic unit cell staggered magnetization resulting from a broken twofold symmetry.  Directional anisotropy, or often referred to as non reciprocal responses, can occur in antiferromagnetic phases in the absence of the Dzyaloshinskii-Moriya interaction or other effects resulting from spin-orbit coupling.

\end{abstract}

\pacs{}

\maketitle

\section{Introduction}

Mechanisms for controlling the flow of excitations, analogous to diodes in electric circuits, have been sought after to create anisotropic devices for magnonic~\cite{Gladii16:93,Schneider08:77,Mruc17:96}, acoustic~\cite{Fleury14:343,Liang10:9,Liang09:103}, and optical~\cite{Fan12:27,Rikken02:89,Roth00:85} applications.~\cite{Tokura18:9}   An example of anisotropic excitations in bulk materials are spin waves in low crystallographic symmetry crystals~\cite{Katsura05:95,Katsura07:98,Takahashi11:8,Git17:119} where magnons propagating in differing directions have dissimilar velocities.  Such excitations have been defined as non reciprocal~\cite{Deak12:327,Szaller13:87} given that the motion in one direction differs from that in the opposite~\cite{Cheong18:3} in the presence of broken time reversal symmetry.~\cite{Onsager31:37}  Other directional anisotropies have been reported in optical measurements where the response depends on the direction of the incident probing beam~\cite{Vinegrad18:5} resulting in contrasting absorption for counter propagating beams.~\cite{Bordacs12:8,Saito08:77,Szaller13:87}  

Directional anisotropic excitations  have been predicted and measured in the presence of relativistic spin-orbit coupling.  For spin-wave excitations, this has been reported in MnSi~\cite{Weber18:97,Sato16:94}, LiFe$_{5}$O$_{8}$~\cite{Iguchi15:92}, CuV$_{2}$O$_{7}$~\cite{Git17:119} where the antisymmetric Dzyaloshinskii-Moriya interaction has been implicated as the origin of the non reciprocal effects under applied magnetic fields.    All of these materials also host a net ferromagnetic moment at low temperatures providing a direct means of coupling the magnetism to a net applied magnetic field.  We find that directional anisotropy of the magnetic fluctuations can occur in the absence of such spin interactions through the application of neutron scattering at high magnetic fields in antiferromagnetic iron based langasite.

Ba$_{3}$NbFe$_{3}$Si$_{2}$O$_{14}$~\cite{Lee10:22,Zhou09:21,Marty10:81,Chaix13:110} (space group \# 150 P321) is structurally single domain chiral with the 6 symmetry elements including the unity operation  $\{1\}$, two 3-fold rotational symmetry axes along the crystallographic $c$ axis $\{3^{+}_{[0,0,1]}, 3^{-}_{[0,0,1]}\}$, three 2-fold axes within the $ab$ plane $\{2_{[x,x,0]}, 2_{[x,0,0]}, 2_{[0,y,0]}\}$ and also time reversal symmetry.  The low temperature (T$_{N}$=27 K) magnetic incommensurate $\vec{q_{0}}=(0,0,\sim 1/7)$ order (with symmetry P3211$'$) removes the 2-fold axes of the underlying crystallographic structure with the exception of those Fe$^{3+}$ spins aligned exactly along a crystallographic 2-fold axes.  We discuss this point further below.  However, the magnetism does preserve the 3-fold symmetry for all Fe$^{3+}$ spins within the $ab$-plane and time reversal symmetry (denoted as 1$'$ in the magnetic space group).~\cite{Marty10:81}   The magnetic structure is based on locally isolated triangles of 120$^{\circ}$ oriented Fe$^{3+}$ ($S$=5/2, $L$=0) spins forming a hexagonal framework in the the $ab$-plane.~\cite{Chaix16:93,Scagnoli13:88}   The Fe$^{3+}$ triangles are stacked along the $c$-axis, linked via two O$^{2-}$ ions giving a helical exchange pathway breaking inversion symmetry.    

Given the underlying nuclear structure, the magnetic spins form a single domain chiral spin pattern below T$_{N}$.~\cite{Quirion17:96, Flack03:86}  The original papers discussing the magnetic structure~\cite{Marty08:101}  defined two terms to characterize the magnetic structure - helicity and chirality. The helicity was used to define the orientation of Fe$^{3+}$ moments on neighboring planes and is fixed through competing symmetric exchange interactions with $J\sim$ 1 meV.  The chirality on an individual Fe$^{3+}$ triangle is set by a weak antisymmetric exchange, allowed through the distortion of the local crystal field environment~\cite{yosida}, with $D$=0.004 meV.~\cite{Stock11:83,Jensen11:84}  The dominant role that symmetric exchange plays in fixing the magnetic handiness makes langasite unique over other chiral magnets such as MnSi, LiFe$_{5}$O$_{8}$, and CuV$_{2}$O$_{7}$ where antisymmetric exchange fixes the underlying chiral magnetic structures.

The spin excitations in magnetically ordered iron based langasite~\cite{Jensen11:84,Loire11:106,Stock11:83} are defined by three modes termed the $w1/w2$ (denoted as $w$ modes hereafter) and $c$ modes following Ref. \onlinecite{Jensen11:84}.  The $c$ modes are the gapless in energy Nambu-Goldstone modes corresponding to a continuous rotation of the spins within the $ab$ plane.  On averaging over the three spins of the individual trimer, there is no net displacement of the magnetic moment in the $ab$ plane and the neutron cross section along (0,0,L) is therefore weak following magnetic structure factors of neutron scattering.  The $c$ mode is chiral reflecting the underlying magnetic structure and has been confirmed by polarized neutron spectroscopy.~\cite{Loire11:106,Moon69:181,Blume63:130}  In contrast, the two $w$ modes are gapped in energy and achiral, corresponding to transverse excitations of the Fe$^{3+}$ spins with net zero out-of-plane displacement.  The $w$ modes correspond to out of plane excitations and hence require a finite energy to overcome any anisotropy that fixes the moments in the crystallographic $ab$ plane. The resulting energy gap value for the $w$ modes is therefore determined by a Dzyaloshinskii-Moriya (DM) anisotropy estimated to be of order $\sim$ 0.004 meV with a direction defined by $\vec{D}_{DM}$=(0,0,1).~\cite{Jensen11:84,Zorko11:107}  

In the absence of an applied magnetic field, time reversal symmetry is maintained and Kramers theorem implies at least a twofold degenerate spectrum, with spin waves excited with momentum transfer $+\vec{Q}$ having the same response as -$\vec{Q}$.  With the application of a magnetic field on a lattice which is non centrosymmetric (and hence $\vec{r} \neq -\vec{r}$), time reversal symmetry is broken and inversion and time reversal operators do not commute with the Hamiltonian and are hence not necessarily conserved.  Therefore, differing spin responses excited with $\pm\vec{Q}$, and hence directional anisotropy, become possible.~\cite{Shelankov92:46}  We now illustrate this effect in iron based langasite.

\section{Experimental Details}

The experiments were carried out on a single crystal grown using the floating zone technique aligned such that reflections of the form (H,0,L) lay within the horizontal scattering plane.  Vertical magnetic field experiments used the Panda (FRM2) and MACS (NIST) cold triple-axis spectrometers with the field applied perpendicular to the chiral crystallographic $c$-axis along (-1,2,0). Experiments with the field aligned within the (H,0,L) horizontal scattering plane were done at RITA2 (PSI) and FLEXX/MultiFLEXX (HZB).  For horizontal field measurements on FLEXX, the field was also rotated 90$^{\circ}$ and aligned along (1,0,0).  We note that (-1,2,0) is parallel to a real space  $[0,y,0]$ 2-fold axis and therefore preserves this crystallographic symmetry. The (100) field orientation is located 30$^{\circ}$ from the real space $[x,x,0]$ and $[x,0,0]$ 2-fold axes, but corresponds to a projection of $\cos(30^{\circ})=0.87$ along a 2-fold axis.  All experiments were done with a fixed final energy defining the energy transfer as $E=E_{i}-E_{f}$. 

\begin{figure}[t]
\includegraphics[width=9.2cm] {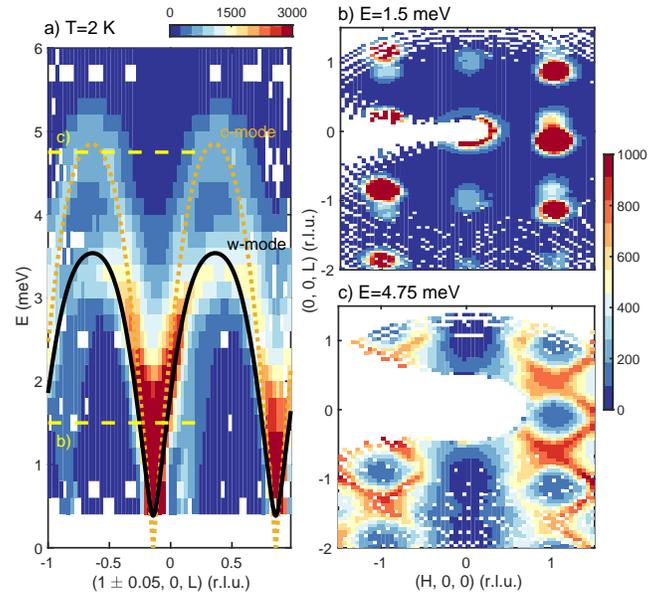}
\caption{\label{figure_1} $(a)$ The spin wave dispersion taken on MACS at 2 K illustrating both $w$ and $c$ modes discussed in the main text with curves from Ref. \onlinecite{Jensen11:84}. $(b-c)$ illustrate constant energy cuts through these two modes.  The intensity along (0,0,L) indicates the polarization of these modes. All data presented in this figure was taken with a fixed E$_{f}$=5.0 meV.}
\end{figure}

\section{Results}

We first discuss the magnetic excitations in zero applied magnetic field.  Figure \ref{figure_1} illustrates the neutron spectroscopic response at $\mu_{0}H=0$ T taken on MACS.  Fig. \ref{figure_1} $(a)$ shows a constant-$Q$ slice along (1$\pm$ 0.05, 0, L) illustrating two distinct magnetic modes.  The lower energy mode corresponds to the $w$ modes while the high-energy mode extending up to $\sim$ 5 meV is the gapless $c$ Goldstone mode following the notation described above.  The solid and dotted curves are the linear spin-wave calculations from Ref. \onlinecite{Jensen11:84} for both the $w$ and $c$ branches. Constant energy slices are displayed in panels Figs. \ref{figure_1} $(b-c)$ at E=1.5 meV slicing through both $w$ and $c$ modes and a higher E=4.75 meV which only crosses through the $c$ mode.  These scans illustrate that the $c$-mode lacks intensity along (0,0,L) (Fig. \ref{figure_1} $c$) showing that, on averaging over the Fe$^{3+}$ spins, no fluctuations perpendicular to the crystallographic $c$ axis occur consistent with this being the gapless Goldstone mode associated with a continuous rotation within the $ab$ plane.  This is in contrast to the lower energy $w$-mode which does show intensity along (0,0,L) (Fig. \ref{figure_1} $b$) indicating a net component with in the $ab$ plane.  These results are consistent with predictions of spin wave theory.~\cite{Jensen11:84}

\begin{figure}[t]
\includegraphics[width=9cm] {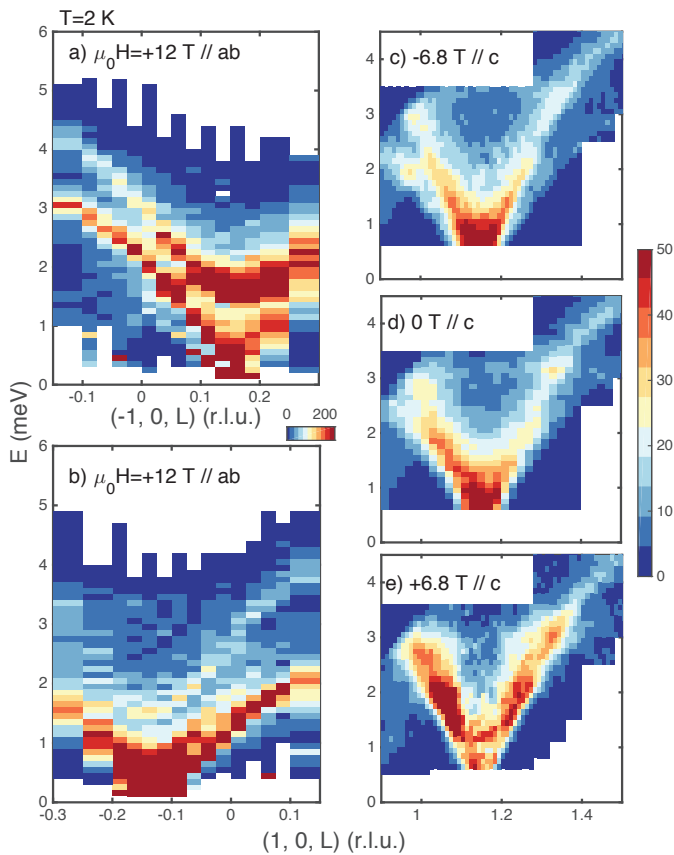}
\caption{\label{field_compare} $(a-b)$ Spin wave dispersions taken with a vertical field aligned within the $ab$ crystallographic plane (fixed final energy of E$_{f}$=5.0 meV).  $(c-e)$ Magnetic dispersion curves taken with a horizontal magnetic field aligned along the crystallographic $c$ axis (fixed final energy of E$_{f}$=3.5 meV).  The white regions were not kinematically reachable given constraints of the horizontal magnetic field.}
\end{figure}

We now discuss the effects of a magnetic field on the spin dynamics.  Figure \ref{field_compare} displays spin waves in langasite for a magnetic field aligned along the $c$ axis and also within the $ab$ plane.  Figures \ref{field_compare} $(a,b)$ show constant $\vec{Q}$ slices taken on Panda with a vertical field (aligned in the crystallographic $ab$ along the (-1,2,0) direction) of 12 T for both Friedal pairs $\vec{Q}$=$\pm$ (1, 0, -1/7) corresponding to opposite total momentum transfers of the incident beam.  The two panels show that the higher energy $c$-mode's dispersion is comparatively weakly affected by the change in sign of $\pm$ $\vec{Q}$ at 12 T.  The $c$-mode excitation is gapped in comparison to zero field (Fig. \ref{figure_1}) owing to the presence of a field induced spin anisotropy.  However, the lower energy $w$-mode shows a contrasting response with (-1, 0, 1/7) displaying a maximum in the dispersion of $\sim$ 3 meV while for the Friedel pair (1, 0, -1/7) the spin waves only reach a maximum of $\sim$ 2 meV. Figures \ref{field_compare} $(a,b)$ illustrates differing spin wave responses in a strong magnetic field when the sign of the total momentum transfer is reversed.  

In Figs. \ref{field_compare} $(c-e)$, we illustrate the dynamics when the magnetic field is aligned along the chiral $c$-axis.   Another means of searching for directionally anisotropic spin waves is to fix $\vec{Q}$ and reverse the sign of the time reversal symmetry breaking magnetic field.  Figs. \ref{field_compare} $(c-e)$ show the effect of reversing the field on the dispersion near the magnetic Bragg peak of $\vec{Q}$=(1, 0, 1/7) illustrating that the low-energy $w$-mode is affected by the field displaying a differing response for $\mu_{0}H$=$\pm$ 6.8 T and 0 T.     Fig. \ref{field_compare} shows that when a magnetic field is applied both within the crystallographic $ab$ plane and along the chiral $c$ axis, an anisotropy is observed in the spin wave response.  We now investigate these two field orientations in detail.

\begin{figure}[t]
\includegraphics[width=9.2cm] {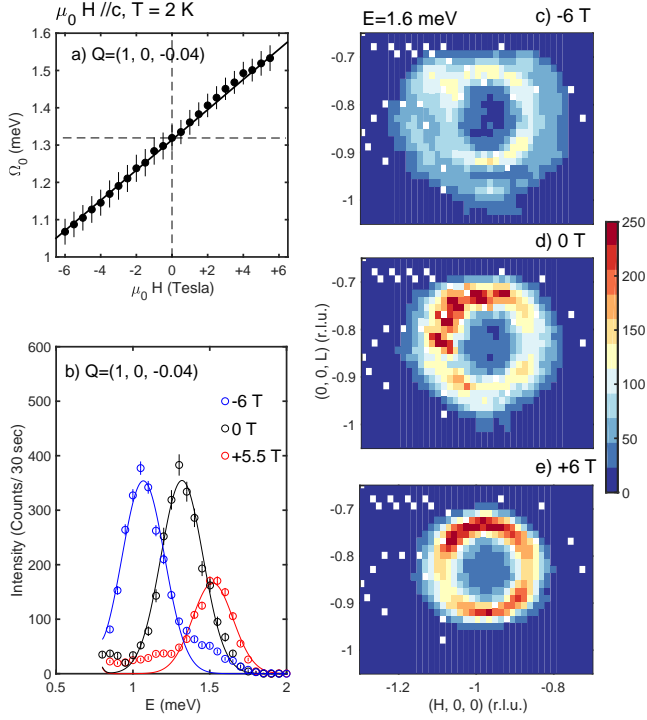}
\caption{\label{field_c} $(a)$ Linear scaling of $\vec{Q}$=(1,0,-0.04) magnon taken using FLEXX (E$_{f}$=2.7 meV). $(b)$ Representative constant momentum scans.  The reduction in intensity at +5.5 T is due to attenuation from the horizontal magnet.   $(c-e)$ display constant E=1.6 meV cuts using the MultiFLEXX detector at 6, 0, and -6T with a fixed final energy of E$_{f}$=2.5 meV.}
\end{figure}

The field dependence of the $w$ mode, when the magnetic field is oriented along the chiral crystallographic $c$-axis, is shown in Fig. \ref{field_c} $(a)$ and $(b)$ by investigating constant $\vec{Q}$=(1,0,-0.04) scans.  The energy position is plotted as a function of applied magnetic field in panel Fig. \ref{field_c} $(a)$ illustrating a linear scaling with field with representative constant momentum scans shown in Fig. \ref{field_c} $(b)$.  The anisotropic in field response of the $w$ spin wave branch is further illustrated through constant E=1.6 meV scans shown in panels Fig. \ref{field_c} $(c-e)$ at $\pm$ 6 and 0 T.   With an applied field of -6 T (Fig. \ref{field_c} $(c)$) a separation between the $w$ and higher velocity $c$ modes is observed.  With increasing fields (Fig. \ref{field_c} $(d,e)$) this circle of intensity fills in with the effect being linear with field. 

\begin{figure}[t]
\includegraphics[width=9.2cm] {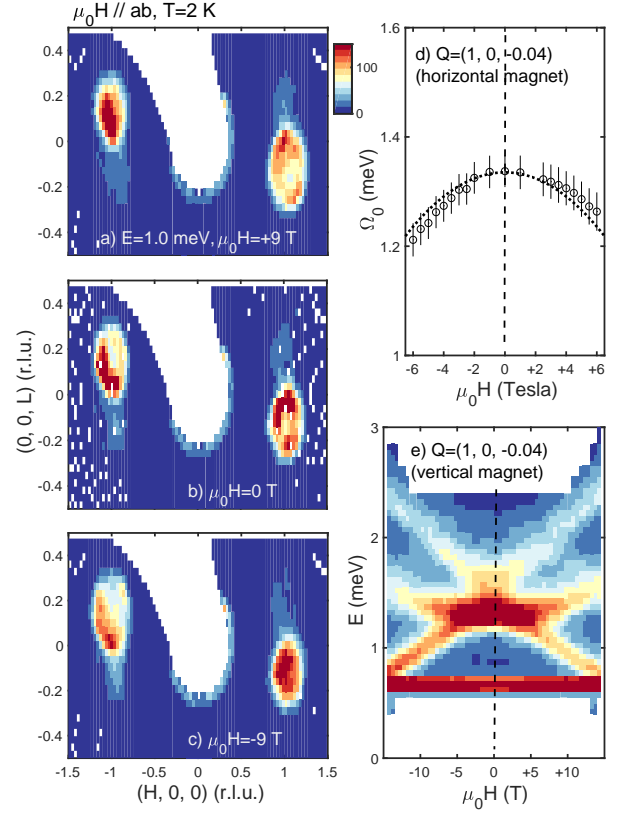}
\caption{\label{field_ab} $(a-c)$ display constant E=1.0 meV constant energy cuts taken on MACS (E$_{f}$=2.5 meV).  $(d$) illustrates the change in energy position of the magnons with $\vec{Q}$=(1,0,-0.04) illustrating a quadratic scaling of position for small fields using a horizontal magnet.  $(e)$ displays constant momentum scans taken with E$_{f}$=2.7 meV on FLEXX showing the isotropic response with magnetic field.  At large applied magnetic fields, the low-energy $w$-mode shows an asymmetric distribution of intensity between the two low-energy modes which are degenerate at zero field.}
\end{figure}

Having observed linear scaling of the $w$-mode with the field oriented along the chiral $c$-axis, we investigate the scaling with the field oriented within the crystallographic $ab$ plane.  Figure \ref{field_ab} shows the response of the lowest energy $w$-mode to a vertical field applied within the crystallographic $ab$-plane.  Figs. \ref{field_ab} $(a-c)$ illustrate constant E=1 meV scans taken using the MACS spectrometer (NIST) which simultaneously measures excitations around the points $\vec{Q}$=$\pm$ (1, 0, -1/7) with magnetic fields of $\mu_{0}H$=0, $\pm$ 9 T.   At an applied field of $\mu_{0}H$=0 T, symmetric spin wave cones are observable at the $\vec{Q}$=$\pm$ (1, 0, -1/7) Friedel pairs.  However, for an applied field of +9T, an apparent contraction of the cone is observed at $\vec{Q}$=(-1, 0, 1/7) while the cone appears increases in diameter at $\vec{Q}$=(1, 0, -1/7).  The opposite response is observed for $\mu_{0}H$=-9T with a contraction of the spin-wave cone at $\vec{Q}$=(1, 0, -1/7) and an increase in diameter at the Friedel pair (-1, 0, 1/7).  

We apply higher resolution scans in Fig. \ref{field_ab} $(d)$ to plot the field dependence of the $w$-mode near $\vec{Q}$=(1,0, -0.04), the same momentum position investigated above for the horizontal field.  These measurements were performed under the same conditions as in Fig. \ref{field_c}
 $(a-b)$, but with the horizontal field rotated by 90$^{\circ}$ within the scattering plane to be aligned along [1,0,0].  Fig. \ref{field_ab} $(d)$ shows a symmetric response with field with the dashed line a fit to $I(\mu_{0}H) \propto (\mu_{0}H)^2$.  Further high resolution measurements using a vertical field allowing larger magnetic fields confirm this symmetric response as shown in Fig. \ref{field_ab} $(e)$.  This figure also shows that 14.5 T is enough to split the degeneracy of the low-energy $w$-mode illustrating their symmetric energy position with $\pm \mu_{0}H$.  However, while the energy dispersion is symmetric with field, the intensity distribution between the two low energy modes is not giving rise to the apparent asymmetric response observed with lower resolution techniques discussed above.  A similar asymmetry in intensity is seen for the $c$-modes for $\pm \vec{Q}$ in Fig. \ref{field_compare} $(a,b)$.  

\section{Discussion and Conclusions}
 
We have observed two different types of directional anisotropy in Ba$_{3}$NbFe$_{3}$Si$_{2}$O$_{14}$.  When the field is applied within the crystallographic $ab$ plane, both the $c$ and $w$ modes display different scattering structure factors for a given $\vec{Q}$ for $\pm$ $\mu_{0}H$.  The energy position of the spin wave branches scales as $\sim H^{2}$ and is directionally isotropic.  When the field is oriented along the crystallographic $c$-axis, the energy position of the low energy $w$ mode displays directional anisotropy with the spin wave energy scaling as $\sim H$. 

To understand these results, we note that the mode that displays directional anisotropy in the energy dispersion is the $w$-mode discussed above.  This mode consists of transverse excitations of the Fe$^{3+}$ moments such $\sum_{i=1-3} \vec{S}_{i,\perp}=0$ (summing over a given isolated triangle Fe$^{3+}$ spins).  The $w$-mode corresponds to a tilting of the plane connecting the Fe$^{3+}$ spins on a given trimer in the P321 unit cell.  No such tilting occurs for the $c$-mode where $\sum_{i=1-3} \vec{S}_{i,||}=0$. Under a magnetic field along the crystallographic $c$-axis, this dynamic tilting would precess around the crystallographic $c$-axis~\cite{Cheon18:98} effectively giving the $w$-mode a helical character absent at zero field.  The size of the effect was predicted based on linear spin wave theory in Ref. \onlinecite{Cheon18:98}  and is in agreement with Fig. \ref{field_c} $(a)$. Because of the globally broken 2-fold symmetry at low-temperatures, spin waves with different directions with differing $\pm \vec{Q}$, are not equivalent.  We note that such a precession does not occur for the gapless $c$-mode and this is consistent with our experiment which finds, with high resolution neutron spectroscopy, the dispersion to be isotropic under $\pm \mu_{0}H$.  Such a precession would also not occur if the magnetic field was perpendicular to the crystallographic $c$-axis as this symmetry axis is preserved in the low temperature magnetic phase.  The importance of the broken 2-fold axis is further highlighted by the observation of directional dichroism for electromagnons.~\cite{Narita16:94} 

The lack of a twofold axis in langasite at low temperatures is inconsistent with an incommensurate magnetic structure where some spins would inevitably align along one of the twofold axes in a bulk crystal.  However, the situation is different when the magnetic structure is truly commensurate ($q_{0}$=1/7).  Then the twofold symmetry is broken by the magnetic structure when the global phase of the helix is chosen such that there is never a magnetic moment aligned exactly parallel or perpendicular to the twofold axis at $z$=1/2 of the sevenfold magnetic unit cell.  It would be interesting to compare the results here against other magnetic systems with definitively incommensurate magnetic structures.

The results here are distinct from MnSi~\cite{Weber18:97,Sato16:94}, LiFe$_{5}$O$_{8}$~\cite{Iguchi15:92}, or Cu$_{2}$V$_{2}$O$_{7}$~\cite{Git17:119,Vasil99:60} where directional anisotropy, or non reciprocal effects, are due to an underlying antisymmetric Dzyaloshinskii-Moriya interaction.   These materials also host a net ferromagnetic moment in the absence of a magnetic field at low temperatures.  Such an effect originates from spin-orbit coupling and indeed has been implicated as being the origin of non reciprocal effects on spin waves in two-dimensional electron gases.~\cite{Perez16:117}  We find that an underlying spin-orbit coupling is not required for the presence of directional anisotropy under a time reversal breaking symmetry field.    In both MnSi and  Cu$_{2}$V$_{2}$O$_{7}$, the non collinear magnetic structures are stabilized by such interactions unlike the case for iron langasite where symmetric exchange stabilizes the helical magnetic structure.  

In summary, we report directional anisotropy for the magnetic excitations in langasite in the magnetically ordered state.  With the field aligned the $c$-axis, directional anisotropy is observed for the spin wave energy response, with the effect scaling linearly with field.  When the field is aligned within the $ab$ plane, the spin wave energies are directionally isotropic, but the intensity is not.  The effect originates from a precession of the dynamic unit cell staggered magnetization and does not originate from antisymmetric exchange as required in other magnetic systems.

\section{Acknowledgements}

We are grateful for funding from the EPSRC, the Carnegie Trust for the Universities of Scotland, the STFC and through the NSF (DMR-09447720).  The work at Rutgers University was supported by the DOE under Grant No. DOE: DE-FG02-07ER46382.   Access to MACS was provided by the Center for High Resolution Neutron Scattering, a partnership between the National Institute of Standards and Technology and the National Science Foundation under Agreement No. DMR-1508249.  We also thankfully acknowledge the financial support from HZB.


%

\newpage

\appendix

\end{document}